\documentclass[fleqn,twoside]{article}
\usepackage{espcrc2}

\usepackage{epsfig}
\usepackage[figuresright]{rotating}
\usepackage{latexsym}
\newcommand{\AmS}{{\protect\the\textfont2
  A\kern-.1667em\lower.5ex\hbox{M}\kern-.125emS}}

\title{Experiences with the multi-level algorithm.}

\author{Pushan Majumdar\address{Max-Planck-Institut f\"{u}r Physik,\\
F\"{o}hringer Ring 6, D-80805 M\"{u}nchen, Germany}
        \thanks{e-mail: pushan@mppmu.mpg.de}}
\begin{document}

\begin{abstract}
Small expectation values are difficult to measure in Monte Carlo 
calculations as they tend to get swamped by noise. 
Recently an algorithm has been proposed by L\"{u}scher and 
Weisz \cite{1} which allows one to measure  expectation values which 
previously could not be measured reliably in Monte Carlo simulations.
We will test our implementation of this algorithm by looking at  
 Polyakov loop correlators and then explore ways of applying it for 
measuring large Wilson loops.
\vspace{1pc}
\end{abstract}

\maketitle

\section{Introduction}

The physical picture of confinement is often given in terms of
formation of a flux tube between a quark and an anti-quark.
This picture is also supported by lattice simulations.

Thinking of the flux tube as a source,
the Wilson loop $W(r,T)$ can be considered as correlation between sources 
of length $r$ propagating for time $T$. 
The theory of this flux tube has been proposed as an effective string 
model \cite{2}.

This is an effective long distance theory and  therefore  to test it, 
 one has to measure physically large Wilson loops.
The problem however is that large Wilson loops have small expectation values. 
That makes it difficult to measure them in simulations. To circumvent 
this problem, recent efforts to measure properties of the flux tube, have
relied on asymmetric lattices \cite{2a}.

Throughout this article,
we will work in 3 dimensions and with SU(2). Even in this restricted 
case, $\langle W \rangle \sim 10^{-8}$ will take about 1 year to measure to an 
accuracy of 1\% with conventional algorithms on a 1Ghz PIII processor.
Here we explore a new algorithm which reduces this time to about a day.

\section{The Algorithm}

The algorithm is based mainly on the factorization property
of the partition function. This makes it possible to obtain
really small expectation values as products of expectation values 
of intermediate orders of magnitude. This works for the Wilson
action to which we restrict ourselves, but can be generalized 
to other actions as long as the factorization property holds.
We do not go into further discussion of the algorithm but just 
present the operational details of our implementation of the 
algorithm. For a detailed discussion we refer the reader to 
\cite{1}.

Suppose we want to measure a Wilson loop
of extent $r$ and $T$ in $x$ and $t$ directions respectively. We choose
the time extent of the lattice to be a multiple of $T$ and
divide the lattice into time slices of thickness $d$. 
At $\beta=4/g_0^2=5$ ($g_0$ is the bare coupling), we found $d=2$ to be 
adequate. Higher $\beta$ might warrant a higher value of $d$.

First we create a 2-link operator (Fig.1A)
\begin{equation}
{\cal T}_{\alpha\beta\gamma\delta}
=U^{\ast}_{1\:\alpha\beta} U_{2\:\gamma\delta}.
\end{equation} 
Next take a product of two \footnote{It can be
any number depending on $T$ and $d$} of these ${\cal T}$'s to get,
\begin{equation}
\widehat{{\cal T}}_{\alpha\beta\gamma\delta}={\cal T}_{1\:\alpha\mu\gamma\nu}
{\cal T}_{2\:\mu\beta\nu\delta}.
\end{equation}
$\langle \widehat{{\cal T}} \rangle$'s are now estimated by an averaging 
procedure in 
which only the time slices are updated holding the space-like links at the 
boundary of the slices fixed. They are then multiplied 
together to form the averaged propagator {\bf T}.
Finally the sources $L_1$ and $L_2$ are calculated and the full Wilson loop
is obtained as (Fig.1B)
\begin{equation}
W=L_{2\:\alpha\gamma}{\bf 
T}_{\alpha\beta\gamma\delta}L^{\ast}_{1\:\beta\delta}.
\end{equation}
\begin{figure}[center]
\vspace*{-4mm}
\mbox{\epsfig{file=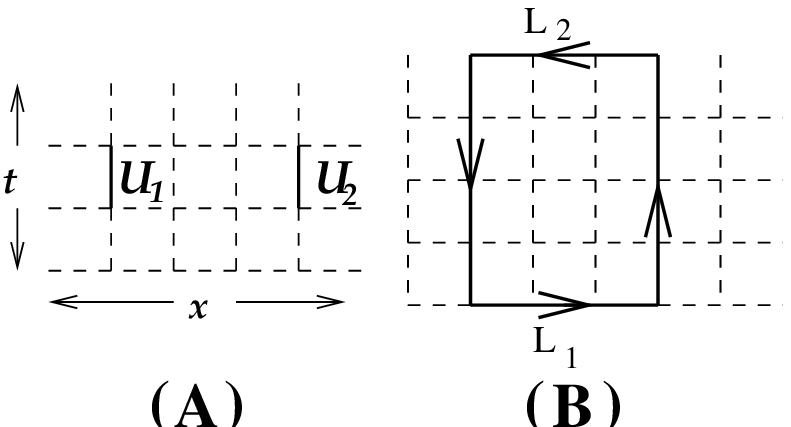,width=5.5truecm,angle=0}}\\
{\centerline{Fig.1. Main steps of the multilevel algorithm}\vspace*{-4mm}}
\end{figure}\vspace*{-4mm}

\section{Optimization}

Let us now try to optimize the parameters of this algorithm.
Probably the most important one  
is the number of sub-lattice updates ($iupd$) and we concentrate on that.
Among the others, we use three over-relaxation steps to 
every heat-bath, choose the basic time slice as two lattice spacings and keep
the level of averaging at one. 

For Polyakov loop correlators ($\langle P^*P \rangle$), the optimal value of 
$iupd$ was estimated in \cite{1} as the one which minimized
$\sqrt{time}\times\langle |P^*P|\rangle$. We follow a slightly different procedure.
The 2-link operator $\widehat{\cal T}$ 
is the main quantity which is affected by the sub-lattice averaging. We propose to 
define the norm of this operator as 
\begin{equation}
N(r)=\sqrt{\sum_{\alpha\beta\gamma\delta\in\{1,2\}} (\langle \widehat{\cal T}(r)
\rangle _{\alpha\beta\gamma\delta})^2}.
\end{equation}
\begin{figure}[htb]
\vspace*{-4mm}
\mbox{\epsfig{file=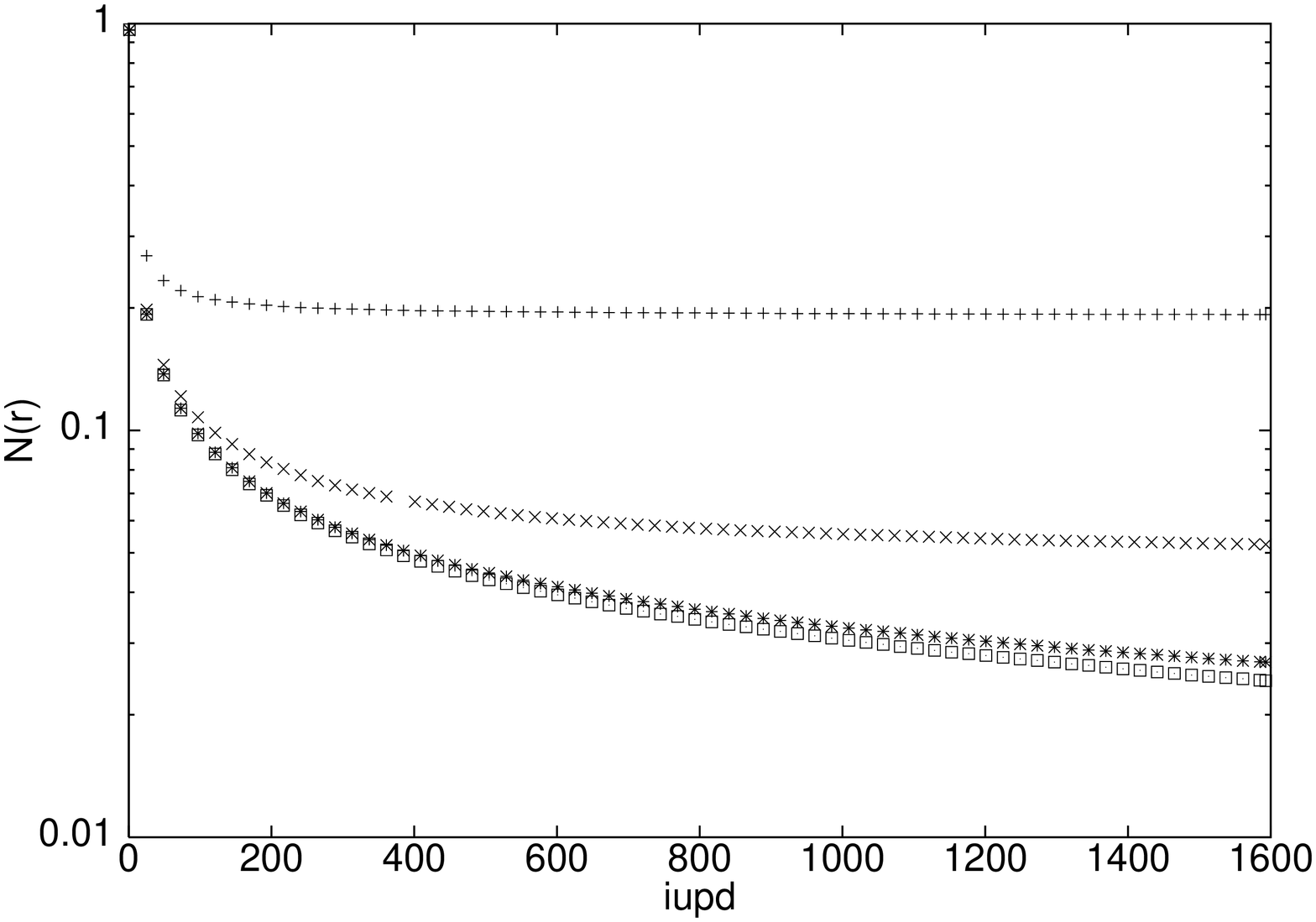,width=4.7truecm,angle=-90}}\\ \\
{\centerline{Fig.2. $N(r)$ versus no. of $iupd$.} 
\centerline{$r$= 2 (+), 4 ($\times $), 6 ($\ast $) and 8 ($\Box $). $\beta=3$.}}
\vspace*{-1cm}
\end{figure}
As seen in Fig.2, $N(r)$ depends quite strongly on $r$. 
Moreover the propagator {\bf T} is a number $O(N(r)^{T/2})$. Thus the optimal value 
of $iupd$ depends not only on $\beta$, but also on the size of the loop itself.
We estimate the optimal value of $iupd$ to be the one which minimizes
$N(r)^{T/2}\times\sqrt{iupd}$. (Note $iupd \propto $ time.)
$N(r)^{T/2}$ is a number $O(\langle |P^*P|\rangle)$, but 
measuring $N(r)$ is cheaper. We checked that both methods give values of $iupd$ 
which are similar. We also find the optimal value of $iupd$ to be reasonably flat and
not too sensitive to the exact definition of $N(r)$. 

For Wilson loops we have to deal in addition with the sources which 
lie entirely in the space slices and whose fluctuation is not
reduced by this algorithm. Note that it is incorrect to apply multi-hit to
any of the spatial links in this algorithm. Fluctuation of the sources
greatly reduce the optimal number of $iupd$ for Wilson loops. 
We do not have any quantitative estimate on how to choose an optimal $iupd$
 for the Wilson loop apriori, but merely point out that if we can estimate the 
order
of magnitude of $\langle W \rangle $ then it is best to choose $iupd$ so that 
each measurement yields a value which is of the same order of magnitude. 
In Fig.3 we plot the error on a $12\times 12$ Wilson loop against $iupd$.
\begin{figure}[center]
\vspace*{-8mm}
\mbox{\epsfig{file=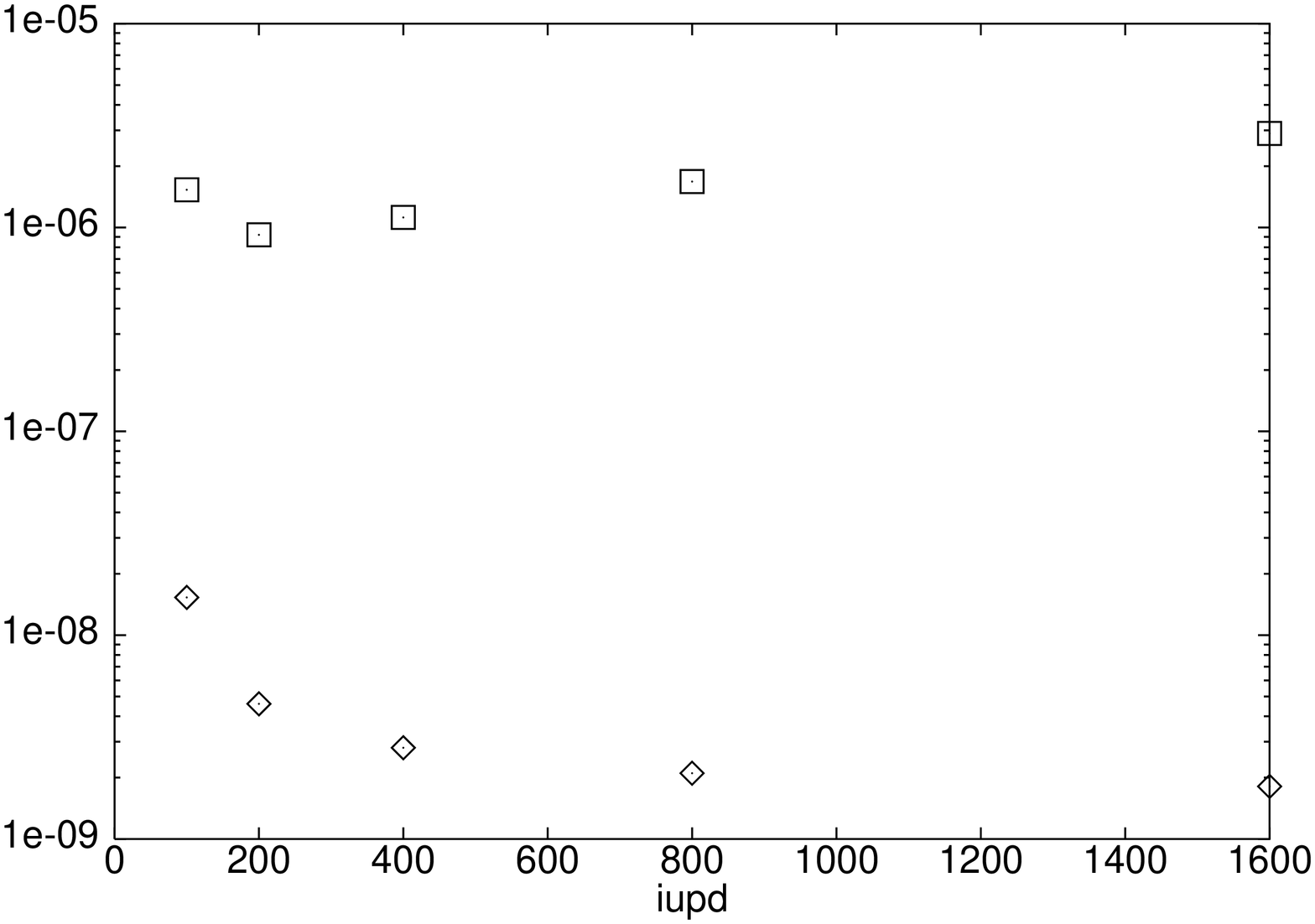,width=4.7truecm,angle=-90}}\\ \\
{ \centerline{Fig.3. error $\times \sqrt{iupd}$ (above) \& error} 
\\ \centerline{(below) vs $iupd$ for $\langle W(12,12) \rangle $at $\beta=5$.}}
\vspace*{-1cm}
\end{figure}
This plot indicates that the optimal value of $iupd$ is $\sim$ 200. In contrast 
the corresponding value for $\langle P^*P \rangle $ is more than 1600.
\section{Results}
\subsection{Polyakov Loop Correlators}
As a test of our program, we compute the 
string tension from $\langle P^*P \rangle $. Neglecting higher 
eigenstates, the static potential between a $\bar{q}q$ pair is 
\begin{equation}
V(r)\approx -\frac{1}{T}\ln\langle P^*(r,T)P(0,T) \rangle.
\end{equation}
In the string picture, for large $r$, $V(r) \approx V_0\: + \alpha r 
\: - c/r $
where $V_0$ is a constant, $\alpha$ is the string tension and $c\:(=\pi/24)$ is 
a universal number.
Next we define the force $F$, and $c$ in terms of $V$ as \\ 
$F(r)=[V(r+1)-V(r-1)]/2$ 
 \hspace{5mm} and \\ $c(r)=[V(r+1)-2V(r)+V(r-1)]r^3/2$. \\We tabulate our 
results below.
\vspace*{-4mm}
\begin{table}[htb]
\caption{$V(r)$, $F(r)$ and $c(r)$ at $\beta=5$. Lattice 24$^3$.}
\label{table:1}
\begin{tabular}{llll}
\hline
 $r$ & $V(r)$ & $F(r)$ & $c(r)$ \\
\hline
 2 & 0.3460(2) & \hspace*{5mm} -&\hspace*{5mm} - \\
 3 & 0.4634(3) & 0.1129(2) &  $-0.1242(8)$\\
 4 & 0.5717(5) & 0.1062(2) &  $-0.131(2)$\\
 5 & 0.6759(8) & 0.1032(3) &  $-0.127(4)$\\
 6 & 0.778(1) & 0.1016(4)  &  $-0.113(9)$\\
 7 & 0.879(2) & 0.1009(5)  &  $-0.08(2)$\\
 8 & 0.980(2) & \hspace*{5mm} - &\hspace*{5mm} - \\
\hline
\end{tabular}
\end{table}
\vspace*{-7mm}
In each measurement, $V(r)$, $F(r)$ and $c(r)$ were obtained from the same 
sample for all $r$ values to take advantage of correlations while computing 
the differences.  The averages and errors were obtained using the jackknife 
method. Finite volume corrections and higher state contributions were 
negligible. 

To get the asymptotic value for $\alpha $ and $c$, we plot $F(r)$ vs 
$1/r^2$. The slope and intercept of this plot gives $c$ and $\alpha$ at 
$r=\infty$. We get $c=0.134(5)$. This is very close to the 
string value of $\pi/24$. The string tension
$a\sqrt{\alpha}=0.313(3)$, compares
very well with Teper's value \cite{3} of $0.3129(20)$.

The above measurement took 10 days on a 1.2 GHz AMD Athlon processor. 

\subsection{Wilson Loops}
Polyakov loops are very good for getting the ground state of the 
flux tube or the string tension, but for the excited states and
states of higher angular momentum, it is more useful to evaluate Wilson 
loops \cite{4}. 

With the view of computing the spectrum of the string, we looked at 
correlation matrices $C_{ij}(r,T)$ between sources $ij$. As a first step 
we define the source 
$s_i= {\cal P}\left( \prod_{k=1}^r(U_k + \alpha_i \sum S_k )\right )$,
with only a single smearing step. $U_k$ is the original link variable,
$\sum S_k$ is the sum over all the space-like staples and ${\cal P}$ 
denotes projection back to SU(2). $\alpha_i$ is the smearing parameter and
takes values $\{0.1, 0.3, 0.5\}$.

The energies are extracted by 
\begin{equation}
E_{\alpha}(r)=-\frac{1}{T_1-T_2}\ln\frac{\lambda_{\alpha}(r,T_1)}
{\lambda_{\alpha}(r,T_2)}.
\end{equation}
 $T_1 > T_2$ and $\lambda_{\alpha}(r,T)$ are the eigenvalues of $C_{ij}$.
We still do not have enough statistics to get the excited states which are  
highly suppressed, and are presently using higher smearing levels and better 
wave functions to extract them. Here we 
present some $T$ dependence for the ground state.
\vspace*{-4mm}
\begin{table}[htb]
\label{Table 2}
\caption{Ground-state for $r=4$ determined from various values of $T_1$ \& 
$T_2$. $\beta=5$.}
\begin{tabular}{c|c|c|c}
\hline
$T_1, T_2$ & 6,4 & 12,4 & 12,6 \\
\hline
$E_0$ & 0.5728(4) & 0.5723(3) & 0.5722(3) \\
\hline
\end{tabular}
\end{table}
\vspace*{-6mm}
The apparent downward trend of the data might be due
to the error \cite{5} introduced for using smaller values of $T_1$ and 
$T_2$.
The energy determined from $\langle P^*P \rangle$, which has 
$T=24$, is 0.5717(5).  

\section{Discussion}
With the multilevel algorithm it is possible to measure $\langle W \rangle 
\sim 10^{-8} $ with 1\% error in about a day, even on a PC.
Even though the algorithm works better for Polyakov loops,
the improvement over multi-hit for $\langle W \rangle < 10^{-6}$ is quite 
significant. 
It is possible to couple this algorithm with other techniques like 
smearing to reduce fluctuation of the sources and study correlations among 
them. For other applications of the algorithm, see \cite{6}.


\begin{thebibliography}{9}
\bibitem{1} M. L\"uscher, P. Weisz, JHEP 0109 (2001) 010.
\bibitem{2} M. L\"uscher, K. Symanzik and P. Weisz,  Nucl. Phys. B173 (1980) 
365.
\bibitem{2a} K. J. Juge, J. Kuti and C. Morningstar, hep-lat/0207004
\bibitem{3} M. J. Teper, Phys. Rev. D59 (1999) 014512. 
\bibitem{4} L. Griffiths,  C. Michael and P.E.L. Rakow, 
Phys. Lett. B129 (1983) 351. 
\bibitem{5} M. L\"uscher, U. Wolff, Nucl. Phys. B339 (1990) 222.
\bibitem{6} Contributions by Slavo Skratochvila and Philip de Forcrand, 
these proceedings.
\end{thebibliography}
\end{document}